\documentclass[conference]{IEEEtran}
\IEEEoverridecommandlockouts

\usepackage{amsmath,amssymb,amsfonts}
\usepackage[backend=bibtex, sorting=none]{biblatex}
\usepackage{graphicx}
\usepackage{subcaption} 
\usepackage{textcomp}
\usepackage{xcolor}
\usepackage{booktabs} 
\usepackage{bm}
\usepackage{amsthm}
\usepackage{algorithm}
\usepackage{algorithmicx}
\usepackage{algpseudocode}
\usepackage{listings}
\usepackage{tikz}
\usepackage{tabularx}
\usepackage{soul}  
\usetikzlibrary{positioning}
\usetikzlibrary{shapes.geometric, arrows.meta, positioning, calc}
\usetikzlibrary{shapes.geometric, arrows.meta, positioning, calc}
\theoremstyle{plain}

\theoremstyle{definition}

\theoremstyle{remark}

\def\BibTeX{{\rm B\kern-.05em{\sc i\kern-.025em b}\kern-.08em
    T\kern-.1667em\lower.7ex\hbox{E}\kern-.125emX}}
\usepackage[acronym]{glossaries}
\newacronym{HVDC}{HVDC}{High Voltage Direct Current}
\newacronym{DT}{DTw}{Digital Twins}
\newacronym{IEC}{IEC}{International Electrotechnical Commission}
\newacronym{CIM}{CIM}{Common Information Model} 
\newacronym{EPRI}{EPRI}{Electric Power Research Institute}
\newacronym{NERC}{NERC}{North American Electric Reliability Corporation}
\newacronym{IED}{IED}{intelligent electronic devices}
\newacronym{LN}{LN}{Logical Nodes}
\newacronym{SCL}{SCL}{Substation Configuration Language}
\newacronym{ACSI}{ACSI}{abstract communication service interfaces}
\newacronym{EMS}{EMS}{Energy Management Systems}
\newacronym{DMS}{DMS}{Distribution Management Systems}
\newacronym{ENTSOE}{ENTSO-E}{European Network of Transmission System Operators for Electricity}
\newacronym{CGMES}{CGMES}{Common Grid Model Exchange Specification}
\newacronym{ESB}{ESB}{Enterprise Service Buses}
\newacronym{UML}{UML}{Unified Modeling Language}
\newacronym{XML}{XML}{eXtensible Markup Language}
\newacronym{SCADA}{SCADA}{Supervision Control and Data Acquisition}
\newacronym{GIS}{GIS}{Geographical Information System}
\newacronym{VSC}{VSC}{Voltage Source Converter}

\definecolor{tagcolor}{rgb}{0,0,0.5}      
\definecolor{attrcolor}{rgb}{0.5,0,0}     
\definecolor{stringcolor}{rgb}{0,0.5,0}   
\definecolor{commentcolor}{rgb}{0.4,0.4,0.4} 
\definecolor{backcolor}{rgb}{0.98,0.98,0.95}

\begin{document}

\title{Use of the Common Information Model \\ (IEC 61970) for Standardized Data Exchange \\ in HVDC Digital Twin Systems \vspace{+10mm}\\

\thanks{The research has been funded by projects Daedalos(Horizon Europe research grant agreement No 101172829). The work of O.Gomis-Bellmunt and E.Prieto-Araujo was supported by the Agència de Gestió d’Ajuts Universitaris i de Recerca (AGAUR) through the ICREA Acadèmia programme, and by the Departament de Recerca i Universitats of the Generalitat de Catalunya. E. Prieto-Araujo is a member of the Serra Húnter Programme.}
}

\author{\IEEEauthorblockN{Ferran Bohigas-Daranas}
\IEEEauthorblockA{\textit{CITCEA} \\
\textit{UPC}\\
Barcelona, Spain \\
0009-0001-6477-6591}
\and
\IEEEauthorblockN{Arkadiusz Burek}
\IEEEauthorblockA{\textit{Hitachi Energy} \\
Sweden \\
 \\
0009-0005-9077-8354}
\and
\IEEEauthorblockN{Eduardo Prieto-Araujo}
\IEEEauthorblockA{\textit{CITCEA} \\
\textit{UPC}\\
Barcelona, Spain \\
0000-0003-4349-5923}
\and
\IEEEauthorblockN{Oriol Gomis-Bellmunt}
\IEEEauthorblockA{\textit{CITCEA} \\
\textit{UPC}\\
Barcelona, Spain \\
0000-0002-9507-8278}
}

\maketitle


\begin{abstract}
\gls{DT} are emerging as transformative tools for \gls{HVDC} transmission systems, enabling real-time monitoring, predictive maintenance, and operational optimization. However, the effectiveness of these virtual replicas fundamentally depends on seamless, reliable data exchange with their physical counterparts and any other support system.

The current landscape of \gls{HVDC} \gls{DT} implementations reveals a critical challenge: the absence of standardized data exchange protocols leads to vendor lock-in, interoperability issues between systems from different vendors. The contribution of this paper is to propose the standardization of \gls{CIM} defined in IEC 61970 standard family, as the data exchange methodology in \gls{HVDC} \gls{DT}s, which is exemplified with the Cigre B4 example, as well as connected to the IEC 61850 standard for real time station control, via the harmonization guidelines from the IEC 62361-102 technical report.

\end{abstract}


\maketitle

  

\section{Introduction}
\label{sec:intro}

\gls{DT}s are emerging as transformative virtual replicas for \gls{HVDC} transmission systems. They enable advanced functionalities including real-time monitoring, operational optimization, and predictive maintenance. Unlike static models, a \gls{DT} provides a virtual representation of physical sites that allows network operators to perform real-time simulations for failure detection and future scenario analysis. The effectiveness of these replicas is fundamentally contingent upon seamless, reliable data exchange between the virtual model and its physical counterpart. In this context, a \gls{DT} acts as an integrated application within a broader ecosystem that includes \gls{SCADA}, \gls{GIS}, and \gls{EMS}.

\gls{HVDC} sites can incorporate equipment from multiple vendors, each employing proprietary data formats and communication protocols. \gls{DT} must interface with diverse systems including \gls{SCADA}, converter control systems, valve monitoring systems, and protection devices, plus other management tools. Without standardization, each interface requires custom development, creating several critical problems:
\begin{itemize}
    \item Integration complexity: Custom interfaces between \gls{DT}s and physical systems require extensive engineering effort, specialized knowledge of proprietary protocols, and ongoing maintenance as systems evolve. This complexity multiplies when \gls{DT}s must integrate data from multiple \gls{HVDC} projects or vendors.
    \item Scalability limitations: As \gls{HVDC} networks expand globally, the ability to replicate and scale \gls{DT} implementations across multiple facilities becomes essential. Proprietary approaches create barriers to scaling, requiring re-engineering for each new plant or technology upgrade.
    \item Lifecycle management risks: \gls{HVDC} facilities operate for decades, often outlasting the original equipment vendors' support lifecycles. Proprietary data exchange methods create long-term dependencies and vulnerabilities when vendors discontinue products or modify their data structures.
\end{itemize}

The economic case for standardization is substantiated by industry evidence, which indicates that data integration  consumes an important part of \gls{DT} implementation budgets \cite{oettl_cost_2023}. Standardized interfaces can dramatically reduce these costs through reusable components, reduced testing requirements, and faster deployment cycles. Furthermore, standardization enables competitive markets for \gls{DT} solutions, preventing monopolistic pricing structures that emerge from vendor lock-in.
From a technical perspective, standardized data exchange enables advanced applications that span multiple facilities or integrate \gls{HVDC} systems with broader grid operations. Multi-terminal \gls{HVDC} systems, \gls{HVDC} grids, and hybrid AC/DC networks require coordinated digital representations that are most effectively achieved through common data semantics and exchange mechanisms.

The \gls{CIM} defined by the \gls{IEC} in \gls{IEC} 61970 standard offers a mature, proven framework particularly well-suited for this standardization. Its semantic comprehensiveness, established interoperability mechanisms, and alignment with broader power system modeling practices make it an ideal foundation for \gls{HVDC} \gls{DT} data exchange. While implementation requires careful attention to \gls{HVDC}-specific requirements and thoughtful architectural design, the benefits, like reduced costs, enhanced interoperability, and future-proofed systems, strongly justify the adoption of \gls{CIM}-based approaches.

As the \gls{HVDC} industry continues its rapid growth and \gls{DT}s become increasingly central to plant operations, the establishment of standardized data exchange methodologies should be recognized as a strategic objective. 

The main contributions of this paper are :
\begin{itemize}
    \item Emphasizing interoperability to facilitate seamless data exchange among the diverse range of applications, including the \gls{DT}, and systems operating within \gls{HVDC} stations and links.    
    \item Proposing the standardization of the \gls{CIM} for \gls{HVDC} \gls{DT}s, specifically within the context of the Common Information Model (IEC 61970/\gls{CIM}), as the most suitable standardized approach to interconnect the different applications and sources which interact with the \gls{DT} applications.
 
\end{itemize}

The remaining sections are structured as follows. Section \ref{sec:interoperability} defines what is interoperability and why is important on commercial control systems. Section \ref{sec:IEC61970basics} describes which are the basic concepts for \gls{IEC} 61970/\gls{CIM}. Section \ref{sec:HVDC_suitability} explains why the \gls{CIM} models are a well-suited option for structuring data in \gls{HVDC} stations \gls{DT}s. Section \ref{sec:CIM-IEC61850} details the interrelation between real-time communications and \gls{CIM} model, by analyzing the interaction between the IEC 61850 and IEC 61970 standards. Section \ref{sec:CigreB4model} exemplifies how the DCS1 link in the Cigre B4 example can be modeled using the Common Information Model. Finally, Section \ref{sec:Conclusions} presents the conclusions of this paper.

\section{Interoperability in monitoring and control systems in Smart Grids}\label{sec:interoperability}

Interoperability is the capability of different hardware and software components, often from various vendors and serving diverse functions, to connect, communicate, and exchange data effectively.

The transition toward the Smart Grid paradigm necessitates a seamless integration of diverse \gls{IED}, sensors, and software platforms across all levels of the electrical infrastructure. Central to this evolution is the concept of \textit{interoperability}. In the context of monitoring and control systems, interoperability is defined as the ability of two or more networks, systems, devices, applications, or components to exchange information and to use the information that has been exchanged to perform required functions \cite{ieee_ieee_2011}. Unlike simple interchangeability, interoperability implies that exchanged data carries sufficient semantic context to be correctly interpreted across heterogeneous systems.

Within electrical asset control and monitoring, two main standards define both semantic models and communication media, IEC 61850 designed for real-time communication between devices inside the sites (station, substation or plant) and IEC 61970/CIM used for exchange information between different applications.

\subsection{Interoperability at the Substation Level: \gls{IEC} 61850}

The \gls{IEC} 61850 standard represents a paradigm shift from traditional protocol-based communication to an object-oriented modeling approach. It provides a comprehensive framework for substation automation that addresses the requirements of high-speed protection, control, and monitoring. 

Key features that enable interoperability within \gls{IEC} 61850 include:
\begin{itemize}
    \item \textbf{Data Modeling:} The standard breaks down physical devices into \gls{LN}, which represent atomic functions (e.g., \textit{XCBR} for a circuit breaker or \textit{PTOC} for time overcurrent protection). This standardizes the data structures regardless of the hardware manufacturer.
    \item \textbf{\gls{SCL},} allows for a vendor-independent exchange of configuration data between system engineering tools. By using \gls{XML}-based files (.SCD, .CID), different devices can be integrated into a single engineering environment.
    \item \textbf{Communication Services:} Beyond the data model, the standard defines abstract communication service interfaces (ACSI) that can be mapped to specific protocols like MMS (Manufacturing Message Specification) or high-speed GOOSE (Generic Object Oriented Substation Event) messaging for real-time peer-to-peer communication.
\end{itemize}

\subsection{The Role of Interoperability in Promoting Free Vendor Competence}

The strategic importance of interoperability extends beyond technical efficiency; it is a fundamental requirement for a competitive and transparent market. Historically, the utility industry was plagued by proprietary "black-box" systems that resulted in vendor lock-in. Once a utility committed to a specific manufacturer's proprietary protocol, the costs of integrating components from a competitor were often prohibitive.

The adoption of open standards like \gls{IEC} 61850 and \gls{CIM} breaks these monopolies by:
\begin{enumerate}
    \item \textbf{Reducing Integration Costs:} Standardized interfaces eliminate the need for expensive custom-made protocol converters and gateways.
    \item \textbf{Fostering Best-of-Breed Selection:} Utilities are empowered to select the most advanced or cost-effective \gls{IED}s and applications for specific functions, knowing they will integrate with existing infrastructure.
    \item \textbf{Stimulating Innovation:} Open standards lower the barrier to entry for smaller, innovative technology firms, as they can develop products that are natively compatible with global grid infrastructures.
    \item \textbf{System Longevity:} By decoupling the software logic from the hardware vendor, utilities can upgrade or replace specific parts of the system without a complete overhaul.
\end{enumerate}

In summary, interoperability acts as the digital grid framework which ensures that technological progress is driven by performance and innovation rather than proprietary barriers.

\section{\gls{CIM}/\gls{IEC} 61970: A Foundation for \gls{HVDC} \gls{DT} Data Exchange}
\label{sec:IEC61970basics}
While \gls{IEC} 61850 focuses on the horizontal and vertical communication within the substation, the \gls{CIM}, standardized under \gls{IEC} 61970 and \gls{IEC} 61968, provides the semantic bridge to the utility enterprise level, defining a common vocabulary and basic ontology \cite{noauthor_ontology_nodate} for the entire power system, including network topology, asset management, and market operations. .

The objectives of the \gls{CIM} models are to enable the utilization of plug-compatible applications and to safeguard the investments made by utilities in these applications. By establishing standardized data interfaces between different software systems, \gls{CIM} enables utilities to integrate new applications from diverse vendors with significantly less effort and errors dedicated to data integration. \gls{CIM} plays a vital role in simplifying data integration and lowering the overall costs and effort associated with connecting disparate systems \cite{gridappsd_introduction_nodate}. 

\subsection{What is the \gls{IEC} 61970 standards family ?}
The \gls{CIM} is an abstract information model designed to represent an electrical network and the various equipment utilized within it.

Originally developed by \gls{EPRI}\cite{epri_common_nodate} and used initially by the \gls{NERC} in the United States, to handle the exchange of data in the transmission level of the power system, due to the reasons mentioned before, now it stands as an open standard, developed and maintained collaboratively by the electric power industry, and has been officially adopted by the \gls{IEC}, under the standard family IEC 61970 series, growing to a much more complex and comprehensive model now used by companies around the world, from the \gls{ENTSOE} in Europe to many utilities in Asia.

The \gls{IEC} 61970 series, developed by the \gls{IEC}, defines the Common Information Model for energy management systems, providing a comprehensive semantic framework for representing power system components, their relationships, and operational data. Originally focused on AC transmission systems,  \gls{CIM} has evolved to include diverse power system elements from distribution to transmission, including support to  \gls{HVDC} systems.

Initially, \gls{CIM} found widespread use in facilitating data exchange for bulk transmission power systems. However, its application is now increasingly extending to include distribution modeling and analysis.
This broadening scope reflects the growing recognition of the need for standardized data management across all voltage levels of the power system. The increasing integration of distributed generation and the deployment of smart grid technologies have amplified the importance of a unified approach to data handling in these evolving networks.

It is vital to recognize that  \gls{CIM} is fundamentally a semantic model \cite{anderson_enabling_2022}\cite{covarrubias_cim_2024}. As such, it provides an abstract and formal depiction of power system objects, their inherent attributes, the relationships that bind them, and the operations that can be performed upon them. \gls{CIM} should not be confused with a database structure or a physical data store. Instead, it functions as a common vocabulary for describing power system data in a manner that is independent of any specific implementation or underlying storage mechanism. 

\subsection{The Role of IEC 61970 Standards}
\begin{figure*}[htpb]
    \centering    \includegraphics[width=0.8\linewidth]{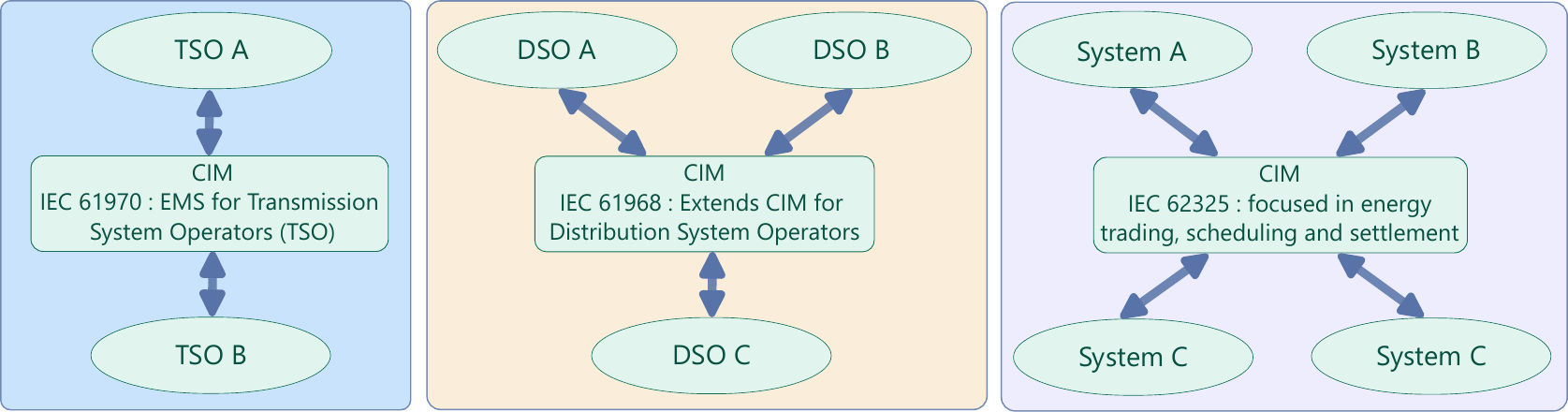}
    \caption{\gls{IEC} 61970 family }
    \label{fig:61970family}
\end{figure*}

The \gls{IEC} 61970 series of standards specifically (Fig. \ref{fig:61970family})\cite{dnv_iec_61970_family} addresses the application program interfaces (APIs) for \gls{EMS}. These standards offer a comprehensive set of guidelines and specifications designed to facilitate the seamless integration of applications developed by various suppliers. Furthermore, they enable the efficient exchange of information with systems located external to the control center environment. 

The \gls{IEC} 61970 series is structured into several key parts to address different aspects of EMS-API and information modeling:
\begin{itemize}
\item \textbf{Part 1:} Outlines the general guidelines and requirements necessary for the application of the EMS-API interface standards.
\item \textbf{Part 2:} Provides a comprehensive glossary of terms used throughout the \gls{IEC} 61970 series.
\item \textbf{Part 3XX:} Focuses specifically on the \gls{CIM}, defining the abstract model and its various components.
\item \textbf{Part 4XX:} Details the Component Interface Specification (CIS), which describes the standard interfaces for information exchange.
\item \textbf{Part 5XX:} Specifies the technology mappings for the CIS, outlining how the abstract interfaces are implemented using specific technologies.
\end{itemize}

The parts of the \gls{IEC} 61970 series most directly relevant to the \gls{CIM} fall within the \gls{IEC} 61970-3XX range. These sections are dedicated to defining the \gls{CIM} base model and other logical views of the power system. The core of the \gls{CIM} is established in \textbf{\gls{IEC} 61970-301}, officially titled "Energy management system application program interface (EMS-API) - Part 301: Common information model (\gls{CIM}) base". The most recent version of this standard is \gls{IEC} 61970-301:2020 (Edition 7.0), which incorporates significant technical enhancements, including the addition of ICCP configuration modeling functionalities. Beyond the base model, other crucial standards within the \gls{IEC} 61970 series extend the \gls{CIM} to cover specialized areas. These include \textbf{IEC 61970-302} (\gls{CIM} Dynamics), which addresses the modeling of dynamic system behavior, \textbf{\gls{IEC} 61970-452} (\gls{CIM} Static Transmission Network Model Profiles), which defines profiles for static network representations to execute state estimation and power flow applications, and \textbf{\gls{IEC} 61970-457} (Dynamics Profile), which specifies a standard interface for exchanging power system dynamic model information needed to support the analysis of the small-signal  and/or transient stability analysis. These extensions demonstrate how the fundamental \gls{CIM} base model is adapted and expanded to address the specific requirements of various functionalities and domains within power systems.

\gls{IEC} 61970-301 is a semantic model that describes the components of a power system at an electrical level and the relationships between each component.


\subsection{\gls{CIM} Models and Interoperability}
\begin{figure}[b]
    \centering
    \includegraphics[width=1\linewidth]{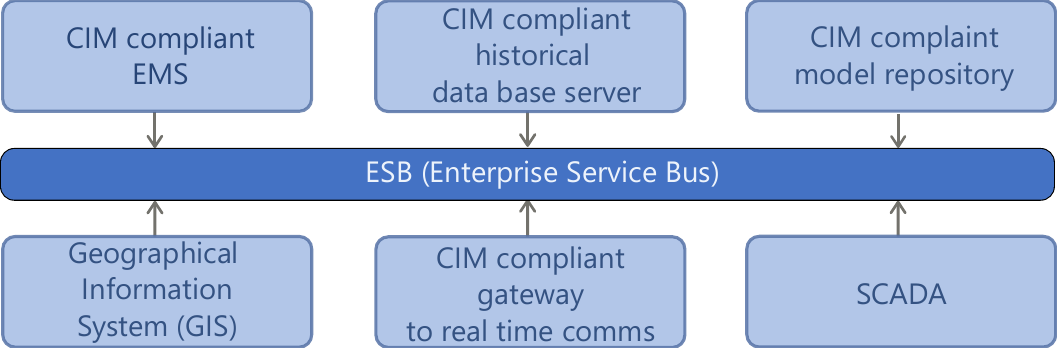}
    \caption{Communication architecture based on \gls{ESB}}
    \label{fig:ESB-CIM}
\end{figure}

A fundamental motivation behind the development and widespread adoption of \gls{CIM} in the power industry is the critical need to achieve seamless interoperability among the applications and systems that are essential for the operation and management of modern power grids. This can be accomplished through the use of \gls{CIM} adapters, specialized software components that translate data from a system's native format into the standardized \gls{CIM} format, and vice versa. Furthermore, \gls{ESB} are often employed as central communication infrastructures (Fig.\ref{fig:ESB-CIM}) that facilitate the routing and exchange of \gls{CIM}-based messages between various applications, thereby simplifying complex integration scenarios. 

In the context of an HVDC Digital Twin, the ESB fulfills a critical architectural role: it decouples the \gls{DT} application layer from the heterogeneous data sources that feed it, including SCADA gateways, protection and control systems, asset management databases, and GIS platforms, by providing a unified, CIM-compliant message exchange layer. Rather than requiring bespoke point-to-point interfaces between each data source and the DT, each system publishes and consumes CIM-formatted messages through the ESB, following a publish-subscribe or request-response pattern. In practical terms, this means that a state change in the physical twin (such as a DC line segment switching out of service or a VSC converter transitioning between control modes) triggers a CIM-encoded event on the ESB that is simultaneously received and processed by the \gls{DT} model repository, the EMS, and any other subscribed application. This architecture also provides a natural boundary for data quality enforcement and semantic validation, ensuring that only well-formed, CIM-compliant data enters the DT's virtual representation of the physical plant.

\subsection{Key Components and Fundamental Concepts of \gls{CIM}}

\begin{figure*}
    \centering
    \includegraphics[width=0.85\linewidth]{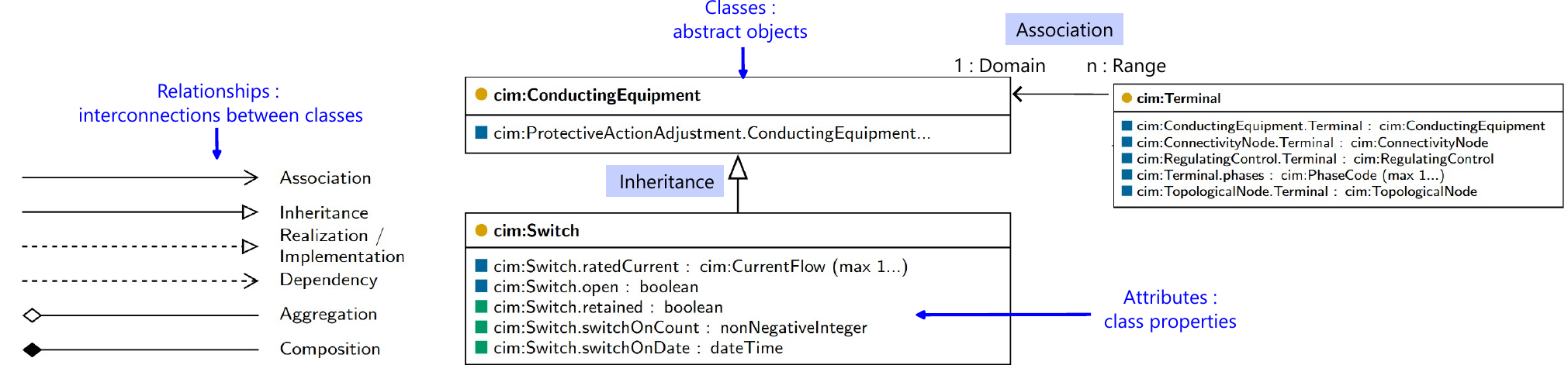}
    \caption{CIM classes and their relationships}
    \label{fig:CIM_theory}
\end{figure*}

The \gls{CIM} is primarily defined and represented using the \gls{UML}, a standardized graphical language that allows for the clear and precise depiction of the structure and behavior of entities within power systems. UML facilitates the definition of fundamental concepts (Fig. \ref{fig:CIM_theory}) such as:
\begin{itemize}
    \item \textbf{classes}, which are abstractions that describe tangible and abstract elements of the power system (for instance, transformers, transmission lines, ...) \cite{uslar_common_2012}, 
    \item \textbf{attributes}, which describe the inherent properties of these classes (like voltage rating, line impedance, ...).
    \item\textbf{relationships}, which illustrate how these classes are interconnected and interact with one another (e.g., a transformer is connected to a transmission line, ...), being main types the associations, aggregations and multiplicities. The \texttt{domain} is the source class of the relationship, while the \texttt{range} is the target class, while the \texttt{multiplicity} defines how many instances of the domain/range class can be related to one instance of the other class, e.g. In the association between Breaker and Terminal, the breaker is the domain, while the terminal is the range with a multiplicity of 2, So 1 breaker is associated with 2 terminals.
\end{itemize}

One of the bases of the \gls{CIM}'s structure is the principle of \textbf{inheritance}, a powerful feature of UML where more specialized classes automatically acquire the properties and behaviors of more general classes. This hierarchical arrangement allows for a more efficient and organized representation of power system components. For example, in the CIM, a \texttt{Breaker} is defined as a specific type of \texttt{ProtectedSwitch}, which in turn is a specialized form of a more general \texttt{Switch}. This inheritance mechanism promotes code reuse and facilitates a more intuitive understanding of the relationships between different power system elements. 

The \gls{CIM} framework is logically organized into several key \textbf{packages}, each serving to group related concepts and functionalities. These packages include:

\begin{itemize}
    \item \texttt{Core} which contains the building block classes common to all applications
    \item \texttt{Wires} which defines the electrical characteristics of various equipment
    \item \texttt{Domain} which specifies the primitive data types used throughout the model
    \item \texttt{Assets} which describes the physical attributes of tangible utility resources;
    \item \texttt{AssetInfo} which holds asset-specific descriptive information.
\end{itemize}

To accurately model the physical connections between components within the power system network, the \gls{CIM} employs the concepts of \textbf{Connectivity Nodes} and \textbf{Terminals}. Physical equipment, such as circuit breakers and power transformers, are represented as \texttt{ConductingEquipment} and possess one or more \texttt{Terminals}. These terminals serve as the points of connection to \texttt{Connectivity Nodes}, which represent locations within the network where the terminals of multiple conducting equipment meet, assumed to have zero impedance. This concept is fundamental to defining the topology of the electrical network within the \gls{CIM}.

Class diagrams show all the attributes and associations of various classes in a particular package. Class specified using \gls{CIM} inherit all the attributes and associations of their parent classes. \gls{CIM} classes also inherit all of the associations of their parent classes to other classes. The highest-level class of objects in \gls{CIM} is called IdentifiedObject. This class is very abstract and only contains attributes used to reference the object either by a user or in software. 


All elements of the power system model are categorized as a \texttt{PowerSystemResource}, which can be any item of equipment (e.g., a switch), an equipment container with multiple pieces of equipment (e.g., a substation or a feeder), or an organizational entity (e.g., a control area). All power system modeling components are then divided into two categories, depending on whether the particular focus is on the electrical connectivity or the physical assets. 

All electrical aspects of equipment are associated with the classes inheriting from the \texttt{Equipment} class in the \gls{IEC} 61970 package. The physical aspects of equipment are associated with the \texttt{Assets} and \texttt{AssetInfo} packages of the 61968 package \cite{anderson_introduction_2023}.

This modular structure not only makes the \gls{CIM} easier to understand and navigate but also facilitates its extension and adaptation for diverse application needs.

The CIM applications involve using descriptive languages such as the \gls{XML} or the Resource Description Framework (RDF), both realized within Service Oriented Architecture (SOA) with corresponding platform support by the middleware.

\subsection{Applications of \gls{CIM} Models in Power Systems}
\gls{CIM} models find extensive application across a diverse range of activities within power system operations and planning. Key areas of application include:
\begin{itemize}
    \item \textbf{\gls{EMS}:} \gls{CIM} serves as the data model for EMS, which are critical for the real-time operation, monitoring, and control of high-voltage transmission networks \cite{santodomingo_iec_2016}.
    \item \textbf{\gls{DMS}:} Extending beyond transmission, \gls{CIM} is increasingly utilized in DMS for the management, analysis, and optimization of power distribution networks, especially with the growth of distributed generation \cite{IEC_61968_2025}.
    \item \textbf{Power System Planning:} \gls{CIM} can play a vital role in various planning activities, including long-term network expansion studies, short-term operational planning involving load flow and contingency analysis, and the integration of new resources like renewable energy. The network planning is going to be covered by the \gls{IEC} 61968-7 standard which is under development.
    \item \textbf{Market Operations:} In deregulated electricity markets, \gls{CIM} is enabling data exchange between different market participants, including utilities, independent power producers, and market operators, supporting functions like energy scheduling and trading. The \gls{IEC} 62325 series of standards further extends \gls{CIM} for energy market communications \cite{IEC_62357_2025}.
    \item \textbf{\gls{CGMES}:} A specific application of \gls{CIM} is CGMES, developed by the European Network of \gls{ENTSOE} \cite{entsoe_entsoe_nodate}. CGMES defines a standardized format based on \gls{CIM} for the exchange of grid models between transmission system operators for various purposes, including system planning and operational studies. The \gls{CGMES} is an \gls{IEC} technical specification (TS) based on the \gls{IEC} \gls{CIM} family of standards, which was developed to meet necessary requirements for TSO data exchanges in the areas of system development and system operation. The \gls{CGMES} can be applied by applications dealing with power system data management, as well as applications the most common analyses (power flow, contingency analysis, short-circuit calculations, capacity allocation, dynamic security assessment)\cite{entsoe_entsoe_nodate}.
    \item \textbf{Asset Management, Outage Management (OMS), and Work Management Systems:} The IEC 61968 series of standards extends the CIM to cover aspects related to the management of utility assets, the handling of planned and unplanned outages, and the coordination of work activities, enabling better integration between these systems and the core operational systems \cite{uslar_common_2012}.
\end{itemize}
These diverse applications underscore the central role of \gls{CIM} in facilitating data integration and interoperability across the power industry.

\section{Technical Suitability of \gls{CIM} model for \gls{HVDC} Applications}
\label{sec:HVDC_suitability}

\gls{CIM} has achieved widespread adoption in transmission system operations globally, supported by major utilities, system operators, and technology vendors. This existing ecosystem provides several advantages for \gls{HVDC} \gls{DT} implementations:
\begin{itemize}
    \item Proven interoperability: \gls{CIM}-based systems have demonstrated successful multi-vendor interoperability in numerous grid operations environments, providing confidence in its technical viability for \gls{HVDC} applications.
    \item Available tools and expertise: The maturity of \gls{CIM} has generated substantial tool support, including model validation utilities, data translators, and software libraries. Professional expertise in \gls{CIM} implementation is widely available, reducing implementation risks.
    \item Integration with grid operations: Many transmission operators already utilize \gls{CIM} for energy management systems. Adopting \gls{CIM} for HVDC \gls{DT}s enables seamless integration with broader grid modeling and operation platforms, supporting holistic system-wide \gls{DT} environments.
    \item Semantic richness: \gls{CIM}'s object-oriented approach provides detailed models for \gls{HVDC} components including converters, DC lines, filters, reactive compensation equipment, and control systems. The standard defines not only the static characteristics of equipment but also dynamic data required for operational \gls{DT}s, including measurements, controls, and status information.
    \item Established information exchange mechanisms: \gls{IEC} 61970 includes standardized profiles and exchange formats, particularly through IEC 61970-301 (\gls{CIM}) and \gls{IEC} 61968-100 (Implementation Profiles). These mechanisms support both bulk data exchange for model synchronization and incremental updates for real-time operations, addressing the diverse data exchange requirements of \gls{DT}s.
    \item Extensibility: The \gls{CIM} framework allows for extensions to accommodate evolving \gls{HVDC} technologies without breaking existing implementations. As \gls{HVDC} systems incorporate new converter topologies, control strategies, or protection schemes, the model can be extended while maintaining backward compatibility.
\end{itemize}

\subsection{Comparison with Alternative Data Exchange Standards}
While several standards exist for data modeling and exchange in energy systems, none offer the same combination of semantic richness, HVDC-specific coverage, and enterprise-level interoperability as CIM. Three alternatives merit specific consideration: OPC Unified Architecture (OPC-UA) and the Asset Administration Shell (AAS) framework defined in IEC 63278.

OPC-UA \cite{OPCfoundation_unified_nodate} is a widely adopted industrial communication framework that provides a flexible, service-oriented architecture with strong real-time data exchange capabilities and native support for secure, encrypted communications \cite{iec_62351_2026}. It has found application in several \gls{DT} implementations in the manufacturing and energy sectors due to its vendor-neutral design and broad tooling support. However, OPC-UA is fundamentally an information transport and modeling framework rather than a domain-specific semantic model. When applied to power systems, it requires the definition of companion specifications to provide meaningful domain semantics, and no mature, standardized companion specification exists today for HVDC systems. This means that, while OPC-UA could serve as a communication layer within an HVDC DT architecture, instead of using IEC 61850, it cannot replace CIM as the semantic foundation for describing HVDC network topology, converter characteristics, or operational state in a way that is interoperable across different utilities and vendors.

The Asset Administration Shell (AAS), standardized under IEC 63278 \cite{iec_63278_2023}, represents an emerging framework from the Industry 4.0 paradigm, designed to provide a standardized digital representation of physical assets throughout their lifecycle. The AAS concept is particularly compelling for asset management use cases, offering fine-grained representation of individual component properties, maintenance records, and lifecycle data. However, the AAS framework currently lacks the network topology modeling capabilities, the established power system ontology \cite{noauthor_ontology_nodate}, and the proven integration with EMS and SCADA environments that CIM provides. Furthermore, the AAS ecosystem for power transmission applications remains immature, with limited standardized submodels for HVDC-specific equipment. CIM, by contrast, has decades of deployment experience in transmission system operations and a well-established ecosystem of compliant tools, validators, and implementation profiles. For these reasons, CIM under the IEC 61970 family represents the most technically mature and operationally proven foundation for data exchange in HVDC Digital Twins, while remaining open to complementary use of AAS for asset lifecycle management functions where its strengths are better suited.

\subsection{Current Limitations and Gaps of CIM for HVDC Applications}
Despite its broad suitability, a critical assessment of the current CIM model reveals several limitations that must be acknowledged when applying it to HVDC Digital Twins. These gaps do not invalidate the proposal to adopt CIM as the data exchange standard, but they highlight areas where the standard requires further development and where implementers must exercise particular care.

The representation of VSC control modes within CIM is functional but incomplete. The current IEC 61970-301 model provides the cim:VsConverter class with attributes for active power control (pPccControl) and reactive power control (qPccControl), supporting common operating modes such as P/Q control and DC voltage regulation. However, the model does not natively capture the full range of advanced control strategies employed in modern VSC-HVDC systems, such as grid-forming control modes including virtual synchronous machine (VSM) operation, droop-based DC voltage sharing in multi-terminal grids, or frequency support functions. These control behaviors are increasingly relevant for HVDC systems operating in weak AC grids or contributing to system stability services, and their absence from the standard CIM model means that implementers must currently resort to CIM extensions or vendor-specific attributes to represent them, undermining the interoperability objective that CIM is intended to serve.

A more significant gap concerns the representation of Modular Multilevel Converter (MMC) technology, which has become the dominant VSC topology in modern HVDC projects. The current CIM model treats the converter largely as a black-box electrical boundary, capturing terminal characteristics and aggregate operational parameters but providing no standardized representation of MMC-specific internal parameters such as sub-module count, arm inductance, arm capacitance, or energy balancing control variables. These parameters are essential for accurate dynamic simulation within a DT environment, particularly for studies involving valve-level interactions, circulating current suppression, or sub-module fault behavior. The IEC 61970-302 dynamics extension partially addresses dynamic converter modeling, but its MMC-specific coverage remains limited compared to what is required for high-fidelity DT applications.

Further limitations exist in the modeling of HVDC protection systems and DC switchgear. As DC circuit breakers become commercially available and DC grid protection schemes are deployed, the CIM model's current representation of DC protection functions is insufficient to describe the fast-acting, highly coordinated protection architectures required in meshed HVDC grids. Similarly, the modeling of DC/DC converters, which are expected to play a role in future HVDC grid interconnections, is not yet standardized within CIM \cite{ivanov_standardized_2024}. These gaps are recognized by the IEC working groups responsible for CIM maintenance \cite{iec_61970_302_2024}, and ongoing standardization efforts, including contributions from different international projects \cite{hvdc-wisehvdc-wise_lib_2025} and CIGRE working groups active in HVDC DT development, are expected to address them in future editions of the standard. In the interim, implementers adopting CIM for HVDC DTs should plan for the use of CIM extension mechanisms, as defined in IEC 61970-301, to accommodate project-specific requirements while maintaining backward compatibility with standard-compliant tools, considering that some of this extensions will become part of the standard itself, if they follow the IEC criteria.

\subsection{Key elements for CIM implementation in HVDC sites}

A \gls{CIM}-based data exchange architecture for HVDC \gls{DT} should incorporate several key elements:
\begin{itemize}
    \item Model repository: A centralized or distributed repository maintaining the authoritative \gls{CIM}-compliant model of HVDC equipment and topology, synchronized with the physical plant configuration.
    \item Data mediation layer: Translation services converting between native equipment protocols and \gls{CIM} formats, handling timing considerations and data quality assurance.
    \item Real-time data services: Mechanisms for publishing operational measurements, status information, and events using CIM-defined data types and temporal semantics appropriate for \gls{DT} applications.
    \item Change management processes: Procedures ensuring that modifications to physical plant configurations are reflected in the \gls{CIM} model with appropriate versioning and change tracking.
\end{itemize}

Applying the above mentioned functionalities should improve the reliability and accuracy of the system models and allow a scalable application development across the power grid \cite{anderson_introduction_2023}


\section{Interrelation between real-time communications and \gls{CIM} model}
\label{sec:CIM-IEC61850}
Together with the \gls{CIM} model, it is also recommended to use a standardized communication protocol to collect the information from the station.  A well-suited and widely adopted option is IEC 61850, given its prevalence in AC and HVDC substation automation, including also a data structure which can be mapped to the \gls{CIM} models \cite{iec_iec_nodate}\cite{berry_introduction_2019}, this will ensure a real time synchronization between the field measurements and signals and the \gls{DT}.

While \gls{CIM} represents a significant step towards achieving interoperability, the power industry also has the challenge of ensuring harmonization between \gls{CIM} and other relevant standards, most notably IEC 61850, which defines the Substation Configuration Language (SCL) used for substation automation. Although both standards aim to standardize data exchange, they were developed independently and have certain differences in their structure and focus. 

Another important point to choose IEC 61850 as the communication protocol for real time data transfer is the substation equipment usually has IEC 61850 available .while CIM models are designed for and primarily deployed in control center and enterprise-level applications, so data collection from the physical twin to the virtual twin will be straightforward.

\subsection{Harmonization via IEC 62361-102}
Achieving enterprise-wide interoperability requires the harmonization of disparate standards, specifically the Common Information Model (CIM) and the IEC 61850 substation automation standard. While IEC 61850 focuses on real-time data exchange within the substation level, CIM (IEC 61970) provides the semantic framework for enterprise-level information exchange. The IEC 62361-102 \cite{iec_iec_nodate,berry_introduction_2019, ling_model_2013, kostic_towards_2003} technical report addresses this gap by defining the mapping and integration methodologies between these two models. This coordinated framework is essential for modern HVDC Digital Twins, as it ensures a cohesive flow of information from physical field measurements at the substation level to the virtual representations in control center applications.

\section{Example of CIM Modeling for a HVDC infrastructure} \label{sec:CigreB4model}

\begin{figure*}[t]
    \centering
    \includegraphics[width=0.75\linewidth]{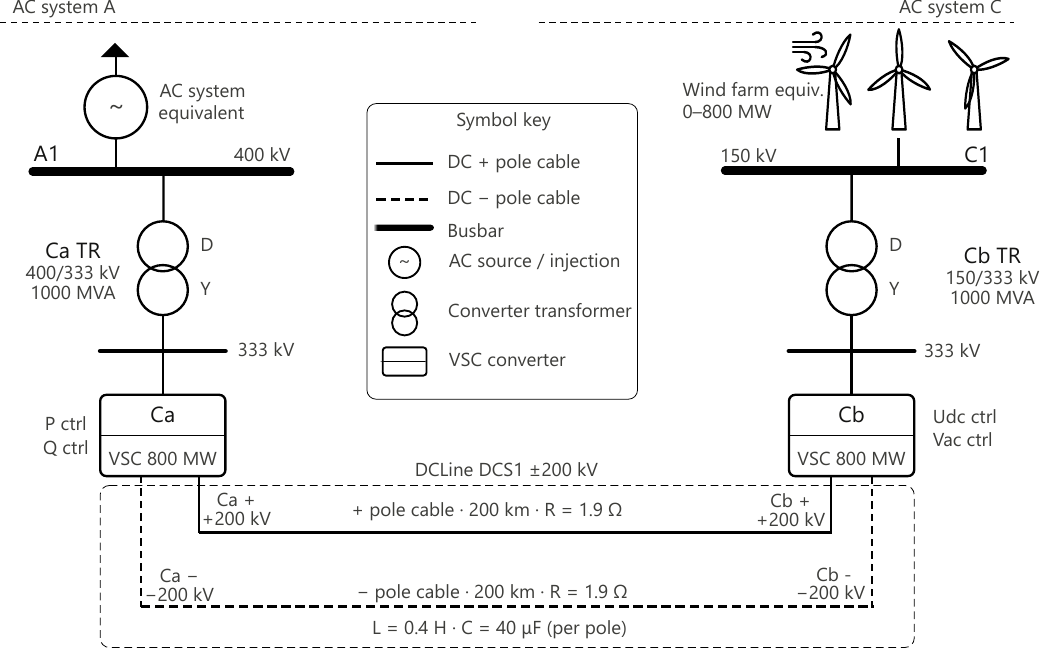}
    \caption{Cigre B4 DCS1 link diagram with CIM elements}
    \label{fig:DCS1_diagram}
\end{figure*}

\begin{figure*}[t]
    \centering
    \includegraphics[width=0.9\linewidth]{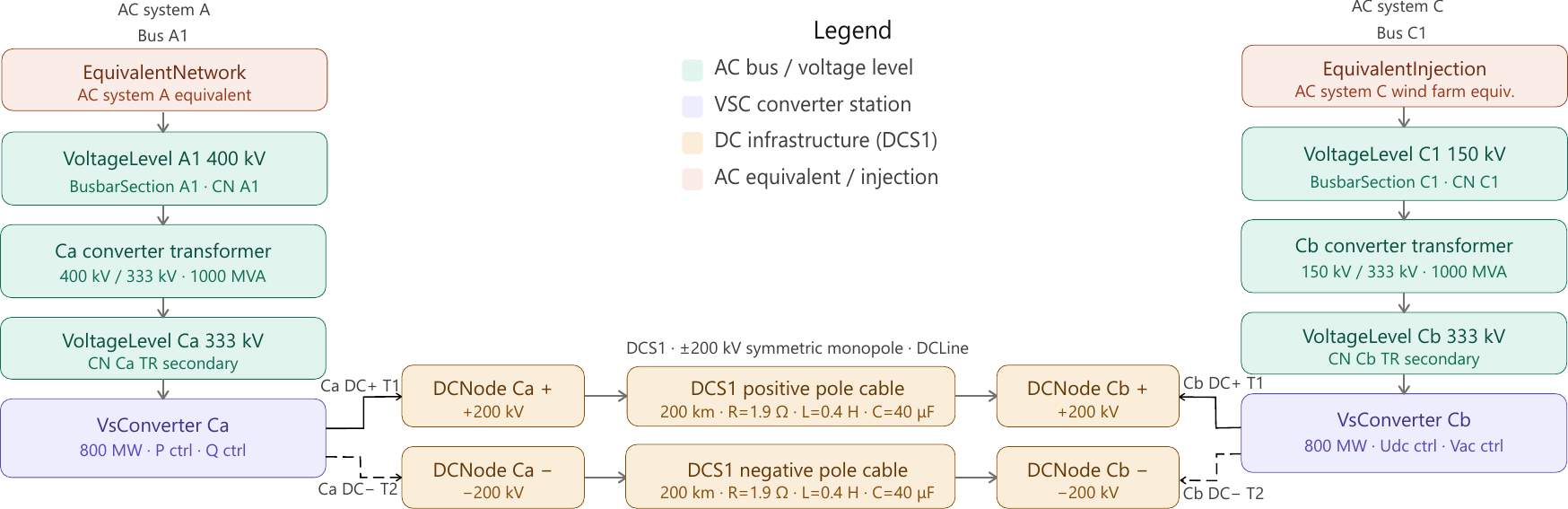}
    \caption{Cigre B4 DCS1 link schema}
    \label{fig:DCS1schema}
\end{figure*}

The following example illustrates the CIGRE B4 DCS1 link (Fig. \ref{fig:DCS1_diagram}), a symmetric monopole link (+/- 200 kV), consisting of two energized conductors, which connected to two \gls{VSC}, one at each end, which are linked to two different AC networks, which are modeled using \gls{CIM} elements, as it can be seen on Fig. \ref{fig:DCS1schema}.

The \gls{VSC} is modeled as \texttt{cim:VsConverter} (Listing \ref{lst:vsc}), a specialization of the general \texttt{cim:ACDCConverter} class, which inherits from \texttt{cim:ConductingEquipment}. The model is logically organized into three distinct sections:

\begin{itemize}
    \item \textbf{AC Side:} This is associated with a \texttt{cim:Terminal} representing the Point of Common Coupling (PCC). It links the converter to the AC grid, typically via a converter transformer (\texttt{cim:PowerTransformer}) connected to an AC bus (\texttt{cim:BusbarSection}).
    \item \textbf{Converter Core:} This section defines the operational setpoints and control modes, such as the DC voltage drop (\texttt{cim:VsConverter.droop}), active power (\texttt{cim:VsConverter.pPccControl}) and reactive power (\texttt{cim:VsConverter.qPccControl}) regulation.
    \item \textbf{DC Side:} Connectivity is managed via the \texttt{cim:ACDCConverterDCTerminal} class. This specialized terminal includes a polarity attribute and links the converter to a \texttt{cim:DCNode}. From the node, the circuit continues through a \texttt{cim:DCTerminal} to a \texttt{cim:DCLineSegment}.
\end{itemize}

Topology is linked to telemetry through the \texttt{cim:ACDCTerminal} base class. All three terminal types (\texttt{Terminal}, \texttt{DCTerminal}, and \texttt{ACDCConverterDCTerminal}) inherit from this base, allowing them to be associated with various \texttt{cim:Measure} types (e.g., AC/DC voltage, active, or reactive power) depending on the terminal's location. Additionally, the \texttt{cim:VsCapabilityCurve} is used to define the physical operating limits imposed by the IGBT ratings and DC bus voltage.

The following snippet presents the \gls{CIM}/\gls{XML} (RDF) serialization of the monopolar VSC-HVDC link, including the AC interface and the DC ground return:

\begin{lstlisting}[float,caption={VSC  Model Serialization}, label={lst:vsc}]
  <cim:VsConverter rdf:ID="vsc-A1">
    <cim:IdentifiedObject.mRID>vsc-A1</cim:IdentifiedObject.mRID>
    <cim:IdentifiedObject.name>Ca-A1 VSC Converter onshore</cim:IdentifiedObject.name>
    <cim:Equipment.EquipmentContainer rdf:resource="#sub-A1"/>
    <cim:ACDCConverter.baseS>1000</cim:ACDCConverter.baseS>
    <cim:ACDCConverter.ratedUdc>200</cim:ACDCConverter.ratedUdc>
    <cim:ACDCConverter.idleLoss>2</cim:ACDCConverter.idleLoss>
    <cim:ACDCConverter.switchingLoss>0.0015</cim:ACDCConverter.switchingLoss>
    <cim:ACDCConverter.resistiveLoss>0.004</cim:ACDCConverter.resistiveLoss>
    <cim:ACDCConverter.numberOfValves>6</cim:ACDCConverter.numberOfValves>
    <cim:ACDCConverter.maxP>800</cim:ACDCConverter.maxP>
    <cim:ACDCConverter.minP>-800</cim:ACDCConverter.minP>
    <cim:ACDCConverter.poleLossP>4</cim:ACDCConverter.poleLossP>
    <cim:VsConverter.maxModulationIndex>1.05</cim:VsConverter.maxModulationIndex>
    <cim:VsConverter.maxValveCurrent>2.5</cim:VsConverter.maxValveCurrent>
    <cim:VsConverter.droop>0.0</cim:VsConverter.droop>
    <cim:VsConverter.droopCompensation>0.0</cim:VsConverter.droopCompensation>
    <cim:VsConverter.pPccControl rdf:resource="#VsPpccControlKind.pPcc"/>
    <cim:VsConverter.qPccControl rdf:resource="#VsQpccControlKind.reactivePcc"/>
  </cim:VsConverter>
\end{lstlisting} 
\section{Conclusion}
\label{sec:Conclusions}
The increasing deployment of Digital Twin technology in HVDC transmission systems creates an urgent need for standardized, interoperable data exchange frameworks that can bridge the diverse ecosystem of vendors, applications, and communication protocols present in modern HVDC installations. This paper has argued that the Common Information Model, standardized under the IEC 61970 family, represents the most technically mature and operationally proven foundation available for this purpose. Through the analysis of CIM's semantic framework, its architectural role within Enterprise Service Bus-based integration environments, and its comparison against alternative standards, including OPC-UA and the Asset Administration Shell, it has been shown that CIM uniquely combines domain-specific power system semantics, established interoperability mechanisms, and a proven deployment ecosystem that alternatives currently cannot match at the enterprise level. The paper has further demonstrated, through the CIM modeling of the CIGRE B4 DCS1 benchmark link, that existing CIM classes (including \texttt{cim:VsConverter}, \texttt{cim:DCLineSegment}, and \texttt{cim:ACDCConverterDCTerminal}) are sufficient to represent the topology, electrical characteristics, and control configuration of a VSC-HVDC system in a standardized, machine-readable format. The proposed integration of CIM with IEC 61850, harmonized through IEC 62361-102, provides a coherent end-to-end data pathway from field-level real-time measurements at the substation to the virtual representation maintained in the DT application layer.

Nevertheless, several limitations of the current CIM standard must be acknowledged. The representation of advanced VSC control modes, including grid-forming operation and droop-based DC voltage sharing in multi-terminal configurations, remains incomplete in IEC 61970-301, requiring implementers to rely on CIM extensions that may compromise cross-vendor interoperability. The absence of standardized MMC-specific parameters (such as sub-module count, arm impedance, and energy balancing variables) limits the fidelity of DT models for dynamic studies involving modern converter topologies. Furthermore, the modeling of DC protection systems and DC switchgear within CIM does not yet reflect the requirements of emerging meshed HVDC grids equipped with DC circuit breakers. From a tooling perspective, while CIM enjoys broad support in AC transmission environments, the availability of validated, HVDC-specific CIM model repositories, conformance testing tools, and data mediation components remains limited, representing a practical barrier to adoption that the industry must collectively address. These limitations do not undermine the fundamental suitability of CIM for HVDC DT data exchange, but they establish a clear agenda for continued standardization work.

Future research and development efforts should therefore focus on several complementary directions. First, the validation of the proposed CIM-based data exchange architecture on a real HVDC installation, encompassing the full data pathway from IEC 61850 field devices through an ESB to a live DT application, is essential to confirm the practical viability of the approach and to identify implementation challenges that are not apparent from a standards analysis alone. Second, the extension of the CIM model to multi-terminal HVDC grids and hybrid AC/DC networks represents a priority standardization objective, as the coordination of multiple converters under a unified semantic model introduces topological and operational complexities beyond those of the point-to-point link demonstrated here. Third, the development of automated or semi-automated tooling for CIM-to-IEC 61850 model mapping, building on the harmonization framework of IEC 62361-102, would significantly reduce the engineering effort required to deploy and maintain HVDC DTs, and represents a practical step toward the plug-and-play interoperability that the standard family aspires to achieve. Finally, as MMC technology continues to dominate new HVDC projects, a dedicated effort to extend IEC 61970-301 and IEC 61970-302 with standardized MMC-specific classes and dynamic parameters, developed in coordination with CIGRE working groups active in this domain, would substantially increase the value of CIM as the data exchange foundation for next-generation HVDC Digital Twins.

\printbibliography

@misc{entsoe_entsoe_nodate,
	title = {{ENTSOE} : {Common} {Information} {Model} ({CIM})},
	author = {ENTSOE},
}

@inproceedings{ling_model_2013,
	title = {Model {Differences} between {IEC} 61970/61968 and {IEC} 61850},
	booktitle = {2013 {Third} {International} {Conference} on {Intelligent} {System} {Design} and {Engineering} {Applications}},
	author = {Ling, Liu and Hongyong, Yu and Xia, Chen},
	month = jan,
	year = {2013},
	pages = {938--941},
}

@inproceedings{kostic_towards_2003,
	address = {Montreal, Que., Canada},
	title = {Towards the formal integration of two upcoming standards: {IEC} 61970 and {IEC} 61850},
	shorttitle = {Towards the formal integration of two upcoming standards},
	booktitle = {Large {Engineering} {Systems} {Conference} on {Power} {Engineering}, 2003},
	publisher = {IEEE},
	author = {Kostic, T. and Preiss, O. and Frei, C.},
	year = {2003},
	pages = {24--29},
}

@article{berry_introduction_2019,
	title = {Introduction to {IEC} 62361-102 {CIM} - 61850 {Harmonization}},
	language = {en},
	author = {Berry, Tom and Electric, Schneider},
	year = {2019},
}

@misc{gridappsd_introduction_nodate,
	title = {Introduction to the {Common} {Information} {Model} — {GridAPPS}-{D} 2021\_05.0 documentation},
	author = {gridappsd},
}

@article{covarrubias_cim_2024,
	title = {{CIM} as a {Framework} for {Smart} {Grid} {Interoperability} {Part}},
	language = {en},
	author = {Covarrubias, Michael},
	year = {2024},
}

@article{anderson_enabling_2022,
	title = {Enabling {Data} {Exchange} and {Data} {Integration} with the {Common} {Information} {Model}},
	language = {en},
	author = {Anderson, Alexander A and Stephan, Eric G and McDermott, Thomas E},
	year = {2022},
}

@article{anderson_introduction_2023,
	title = {An {Introduction} for {Power} {Systems} {Engineers} and {Application} {Developers} – {CIM17v40}},
	language = {en},
	author = {Anderson, Alexander A and McDermott, Thomas E and Stephan, Eric G},
	year = {2023},
}

@misc{noauthor_ontology_nodate,
	title = {Ontology {Documentation}},
	url ={https://ontology.tno.nl/IEC_CIM/},

}

@book{uslar_common_2012,
	address = {Berlin, Heidelberg},
	series = {Power {Systems}},
	title = {The {Common} {Information} {Model} {CIM}: {IEC} 61968/61970 and 62325 - {A} practical introduction to the {CIM}},
	volume = {66},
	copyright = {https://www.springernature.com/gp/researchers/text-and-data-mining},
	shorttitle = {The {Common} {Information} {Model} {CIM}},
	language = {en},
	publisher = {Springer},
	author = {Uslar, Mathias and Specht, Michael and Rohjans, Sebastian and Trefke, Jörn and Vasquez Gonzalez, Jose Manuel},
	year = {2012},
}

@misc{hvdc-wisehvdc-wise_lib_2025,
	title = {{HVDC}-{WISE}/{HVDC}-{Wise}\_lib},
	copyright = {EUPL-1.2},
	publisher = {HVDC-WISE},
	month = apr,
	year = {2025},
	note = {original-date: 2023-10-16T17:08:13Z},
}

@misc{dnv_iec_61970_family,
	title = {CIM training course by DNV},
    author ={DNV},
	publisher = {DNV},
	month = apr,
	year = {2024},
}

@misc{IEC_62357_2025,
	title = {{IEC} {TR} 62357-1:2016},
    author = {International Electrotechnical Commission (IEC)},
	shorttitle = {{IEC} {TR} 62357-1},
}

@misc{IEC_61968_2025,
	title = {{IEC} 61968-11:2013},
    author = {International Electrotechnical Commission (IEC)},
	shorttitle = {{IEC} 61968-11},
}

@incollection{santodomingo_iec_2016,
	title = {{IEC} 61970 for {Energy} {Management} {System} {Integration}},
	copyright = {Copyright © 2016 John Wiley \& Sons, Ltd},
	language = {en},
	booktitle = {Smart {Grid} {Handbook}},
	publisher = {John Wiley \& Sons, Ltd},
	author = {Santodomingo, Rafael and Uslar, Mathias and Specht, Michael and Rohjans, Sebastian and Taylor, Gareth and Pantea, Stefan and Bradley, Martin and McMorran, Alan},
	year = {2016},
}

@article{oettl_cost_2023,
	series = {16th {CIRP} {Conference} on {Intelligent} {Computation} in {Manufacturing} {Engineering}},
	title = {Cost estimation approach of a digital twin implementation in industry},
	volume = {118},
	issn = {2212-8271},
	journal = {Procedia CIRP},
	author = {Oettl, Fabio and Eckart, Leonard and Schilp, Johannes},
	month = jan,
	year = {2023},
}

@article{ieee_ieee_2011,
	title = {{IEEE} {Guide} for {Smart} {Grid} {Interoperability} of {Energy} {Technology} and {Information} {Technology} {Operation} with the {Electric} {Power} {System} ({EPS}), {End}-{Use} {Applications}, and {Loads}},
	doi = {10.1109/IEEESTD.2011.6018239},
	journal = {IEEE Std 2030-2011},
	author = {IEEE},
	month = sep,
	year = {2011},
}

@misc{epri_common_nodate,
	title = {Common {Information} {Model} {Primer}: {Eleventh} {Edition}},
	author = {EPRI},
}

@misc{iec_iec_nodate,
	title = {{IEC} {TS} 62361-102:2018},
	shorttitle = {{IEC} {TS} 62361-102},
    author = {International Electrotechnical Commission (IEC)},	
    author = {IEC},
}

@misc{OPCfoundation_unified_nodate,
	title = {Unified {Architecture} - {Landingpage}},
    author ={OPC foundation},
    url = {https://opcfoundation.org/about/opc-technologies/opc-ua/},
	journal = {OPC Foundation},
}

@misc{iec_62351_2026,
	title = {{IEC} 62351:2026 {SER}},
    author = {International Electrotechnical Commission (IEC)},	shorttitle = {{IEC} 62351},
}

@misc{iec_63278_2023,
	title = {{IEC} 63278-1:2023},
    author = {International Electrotechnical Commission (IEC)},	shorttitle = {{IEC} 63278-1},
}

@misc{iec_61970_302_2024,
	title = {{IEC} 61970-302:2024},
    author = {International Electrotechnical Commission (IEC)},	shorttitle = {{IEC} 61970-302},
}

@inproceedings{ivanov_standardized_2024,
	address = {Dubrovnik, Croatia},
	title = {Standardized {Model} {Exchange} of {HVDC} {Equipment} for {RMS} and {EMT} {Simulations}},
	language = {en},
	booktitle = {2024 {IEEE} {PES} {Innovative} {Smart} {Grid} {Technologies} {Europe} ({ISGT} {EUROPE})},
	publisher = {IEEE},
	author = {Ivanov, Chavdar and Tishenin, Georgii and Lanzarotto, Damiano and Morel, Florent and Monti, Antonello},
	month = oct,
	year = {2024},
	pages = {1--5},
}

\end{document}